\documentstyle[twoside,fleqn,espcrc2,epsf]{article}

\newcommand{\be}{\begin{equation}}
\newcommand{\ee}{\end{equation}}
\newcommand{\bea}{\begin{eqnarray}}
\newcommand{\eea}{\end{eqnarray}}

\newcommand{\bc}{\begin{center}}
\newcommand{\ec}{\end{center}}

\title{First Signs for String Breaking in Two-Flavor QCD}

\author{\frenchspacing
Carleton DeTar \address{Physics Department, University of Utah,
    Salt Lake City, UT 84112-0830, USA},
Urs Heller \address{SCRI, Florida State University,
     Tallahassee, FL 32306-4130, USA},
Pierre Lacock$^{\rm{a}}$\thanks{Talk presented by P. Lacock}
\nonfrenchspacing}

\begin{document}

\begin{abstract}
We have been examining the phenomenon of string breaking in QCD with
two flavors of dynamical staggered quarks.  We construct a transfer
matrix from a combination of ``string'' and ``two-meson'' channels.
Preliminary results with low statistics show the expected signs of
string breaking.
\end{abstract}

\maketitle

\section{INTRODUCTION}

During the past few years there has been renewed interest in
the important phenomenon of string breaking (SB), which is predicted
by QCD, but which lattice QCD simulations for 
a long time have failed to show conclusively.

String breaking is observed as a leveling off of the static
quark-antiquark potential at large separation.  The potential is the
separation-dependent ground-state eigenvalue of the QCD hamiltonian in
the presence of the static quark-antiquark pair.  Traditionally, this
eigenvalue was sought in the Wilson loop observable, which is the
expectation value of the transfer matrix on a ``string'' state with
the string of color flux connecting the static quark and antiquark.
Experience has shown, not surprisingly, that the string state ($S$) is
a very poor variational ansatz for a state that looks more like two
static-light mesons.  Including an admixture of a two-meson component
($M$) should help
\cite{michael,detar,drummond,wittig,knechtli,trottier,stewart,pennanen}.

In the following we report on preliminary results that have
been obtained using the relevant string-string and
string-meson operators for QCD.

\section{STUDYING STRING BREAKING ON THE LATTICE}

Our conventional Wilson loop is computed with APE smearing of the
space-like gauge links.  In hamiltonian language the expectation value
of this operator is the transfer matrix $G_{SS}(R,T)$ between an
initial and final state $S$ consisting of a static quark-antiquark
pair separated by a fat string of color flux.  We enlarge the space by
including a meson-antimeson state $M$ with an extra light quark in the
vicinity of the static antiquark and an extra light antiquark in the
vicinity of the static quark.  Thus we also compute the additional transfer
matrix elements $G_{MM}(R,T)$, $G_{MS}(R,T)$ and $G_{SM}(R,T)$.  They
are diagrammed in Figures 1 and 2.

\vspace{0.2cm}

\begin{figure}[htb]
\vspace{-0.9cm}
\epsfxsize=7.5cm\epsfbox{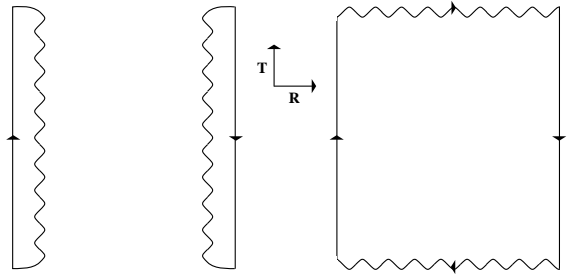}
\vskip -1cm
\hskip -6mm
 \caption{ The static-light meson-antimeson pair contribution
to the full QCD propagator. The wiggly lines denote
the light quark propagator. Shown are the `direct' and 
`exchange' terms respectively.
\label{D1}}
\vspace{-0.9cm}
\end{figure}

\vspace{0.3cm}

In principle, both channels couple to a common set of eigenvalues.  At
large $T$ we expect to reach the ground state, defined by the largest
generalized eigenvalue 
\be
   G(R,T+1)u(R,T) = \lambda(R,T) G(R,T)u(R,T)
\ee
The potential is then given by $V(R) = -\lim_{T\rightarrow
\infty}\log|\lambda(R,T)|$.  The vector $u(R,T)$ defines the
variationally optimum admixture of $S$ and $M$ with the largest
overlap with the ground state.  

A key unitarity condition is that the eigenvalues and eigenstates approach a
constant with increasing $T$.  This condition together with a
demonstration of a smooth transition with increasing $R$ between a
string-dominated state and two-meson-dominated state constitutes a
true test of string breaking and should distinguish a quenched
calculation from a proper calculation with dynamical quarks.

\section{NUMERICAL METHOD}

To maximize statistics we generate ``all-to-all'' propagators for the
light quark, using a Gaussian random source method.  Results reported
here are based on 15 such sources per gauge configuration, but we
plan to increase this number\cite{heller}.

\vspace{0.5cm}

\begin{figure}[hb]
\vspace{-0.9cm}
\epsfxsize=7.5cm\epsfbox{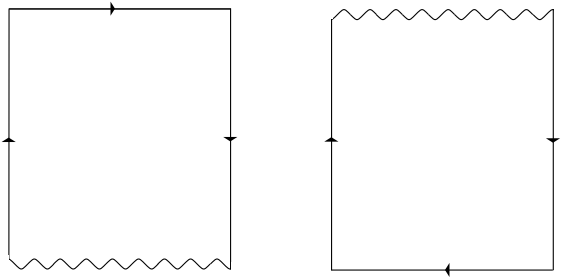}
\vskip -1cm
\hskip -6mm
 \caption{ The string-meson transition correlator $G_{SM}$
(and its hermitian conjugate $G_{MS}$).
The wiggly line again denotes the light quark propagator.
\label{E1}}
\vspace{-0.9cm}
\end{figure}

\vspace{0.2cm}

When constructing operators involving both staggered fermions and
gauge links, one has to pay careful attention to staggered fermion
phases.  A consistent treatment results from interpreting the static
quark as an infinitely massive staggered quark.  For example, a
hopping parameter expansion around an on-axis $R \times T$ rectangular
path gives, in addition to the Wilson-loop gauge-link product, a net
phase factor $(-1)^{RT} \times (-1)^{R+T}$, 
independent of the Dirac phase conventions.  We use a similar
construction to get the phases for the nonclosed gauge link products
in the diagrams of Figures 1 and 2.  In that case the Dirac phase
convention for the gauge link products must be consistent with that of
the light quark. A peculiar consequence of this construction is that
the transition matrix elements must vanish for off-axis displacements
$\vec R$ that have more than one odd Cartesian-displacement component.

\section{LATTICE SIMULATION}

%

We use a set of stored configurations of size $20^3 \times 24$
at $\beta$=5.415 generated with two flavors of dynamical staggered
fermions of mass $ma = 0.0125$. This set gives a ratio $m_{\pi}/m_{\rho}
\approx 0.358$ and has a lattice spacing of about 0.17 fm, which gives a
spatial lattice size of $\approx$ 3.4 fm, and a temporal
size of $\approx$ 4 fm.

For the light fermions in the static-light mesons we use the same
parameters as for the dynamical fermions. With the choice of 15
random sources we have currently analyzed about 70 out of the 200
archived configurations.

\begin{figure}[htb]
\vspace{-0.9cm}
\epsfxsize=7.5cm\epsfbox{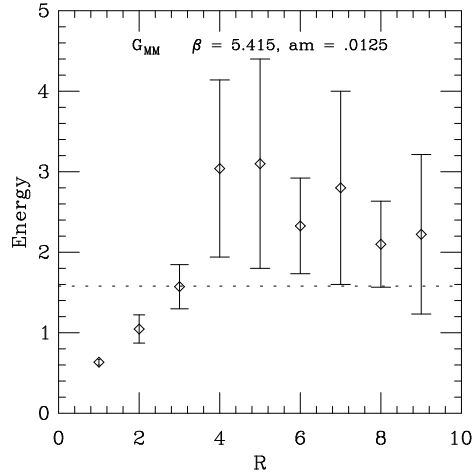}
\vskip -19mm
\hskip -9mm
 \caption{ The behavior of the two-meson to two-meson matrix element $G_{MM}$
as a function
of the spatial separation $R$. Results were obtained from fits
to the data as a function of $T$ for $fixed$ $R$.
\label{A1}}
\vspace{-0.9cm}
\end{figure}

\section{RESULTS}

In Figure 3 we show the results for the two-meson to two-meson
transition $G_{MM}$.  For short distances the operator behaves like
the Wilson loop, with a linearly rising value, while for distances
greater than 3 in lattice units it starts to level off. The dashed
line is twice the mass of the heavy-light meson.

The results for the transition matrix \hyphenation{element} element
$G_{MS}$ shown in Figure 4 are similar within large errors, with the
energy leveling off at $Ra \approx 5$ in lattice units.

In Figure 5 we show the combined results --- \break including those for the
pure Wilson loop correlator. The string-like
behavior of the $G_{MM}$ and, especially, of the $G_{MS}$ correlators
at short distances is clearly visible.

\begin{figure}[htb]
\vspace{-0.9cm}
\epsfxsize=8.0cm\epsfbox{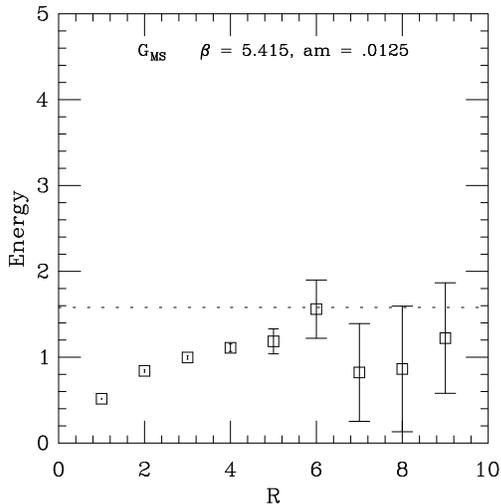}
\vskip -18mm
\hskip -3mm
 \caption{ The same as in Fig. 2, but for the $G_{MS}$ operator.
\label{A2}}
\vspace{-0.9cm}
\end{figure}

\vspace{0.2cm}

\section{CONCLUSIONS}

Our present results with low statistics show a behavior expected with
string breaking.  The cross-over region is found to be at a distance
of $Ra = 5-6$, or about $0.8 - 1.1$ fm.  However, to demonstrate
string breaking convincingly, one must find a smooth transition with
increasing $R$ between a string-dominated state and
two-meson-dominated state and demonstrate that the eigenvalues and
eigenvectors of the transfer matrix approach a constant at large $T$.
Work in this direction is currently underway.

We thank our colleagues of the MILC Collaboration for their help.
This work is supported by the US National Science Foundation and
Department of Energy and used computer resources at the San Diego
Supercomputer Center (NPACI) and the University of Utah (CHPC).


\begin{figure}[htb]
\vspace{-0.9cm}
\epsfxsize=8.0cm\epsfbox{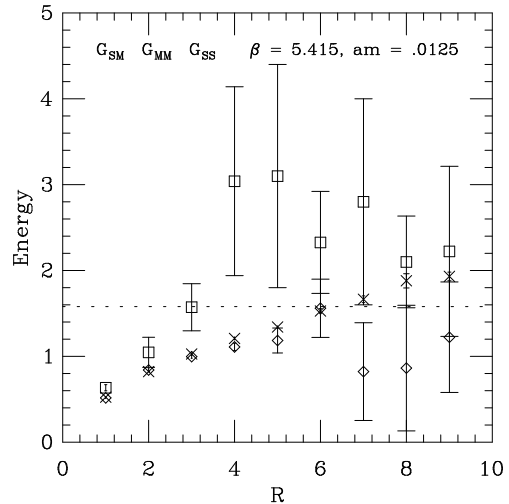}
\vskip -18mm
\hskip -6mm
 \caption{ Combining the results for the $G_{MM}$ ($\Box$),
$G_{MS}$ ($\diamond$)  and Wilson loop
($G_{SS}$) ($\times$) correlators.
\label{B1}}
\vspace{-0.9cm}
\end{figure}


\begin{thebibliography}{9}
\bibitem{michael} C. Michael, Phys. Lett B 283 (1992) 103.
\bibitem{detar}C.~DeTar, O.~Kaczmarek, F.~Karsch and E.~Laermann,
Phys.\ Rev.\ D59 (1999) 031501.
\bibitem{drummond} I. Drummond, Nucl. Phys. B (Proc. Suppl.) 73 (1999) 596.
\bibitem{wittig} O. Philipsen and H. Wittig, 
Nucl. Phys. B (Proc. Suppl.) 73 (1999) 706.
\bibitem{knechtli} F. Knechtli and R. Sommer, these proceedings and
Nucl. Phys. B (Proc. Suppl.) 73 (1999) 584.
\bibitem{trottier} H. Trottier, Nucl. Phys. B (Proc. Suppl.) 73 (1999) 930.
\bibitem{stewart} C. Stewart and R. Koniuk, Nucl. Phys. B (Proc. Suppl.) 73 (1999) 599.
\bibitem{pennanen} P. Pennanen, these proceedings.
\bibitem{heller} C. DeTar, U. Heller and P. Lacock, work in progress.
\end{thebibliography}
\end{document}